\documentclass[prb,twocolumn,showpacs]{revtex4}
\usepackage{array,amsmath,amssymb,graphicx}

\begin{document}
\title{A heuristic  approach to the weakly interacting Bose gas}

\author{Thomas Nattermann}

\affiliation{Institut f\"ur Theoretische
Physik, Universit\"at zu K\"oln,
Z\"ulpicher Str. 77, 50937 K\"oln,
Germany\footnote{e-mail:
natter@thp.uni-koeln.de}}

\date{\today}

\begin{abstract}
Some thermodynamic properties of weakly
interacting Bose systems  are derived from
dimensional and heuristic arguments and
thermodynamic relations, without resorting
to statistical mechanics.

\end{abstract}
\pacs{}

\maketitle
%%%%%%%%%%%%%%%%%%%%%%%%%%%%%%%%%%%%%%%%%%%%%%%%%%%%%%%%%%%%%%%%

\section{Introduction.}

In a previous article \cite{Na_05} I have
derived the thermodynamic properties of
ideal quantum gases solely from
dimensional arguments, the Pauli principle
and thermodynamic relations, without
resorting to statistical mechanics. In the
present publication I will extend this
approach to weakly interacting bosons.

Starting point is the Hamiltonian of
interacting bosons in $d$ dimensions
\begin{equation}
{\cal H} = \int d^dx\Psi^{\dag}\big
(-\frac{\hbar^2}{2m}{\nabla}^2 +
\frac{1}{2}v_0\Psi^{\dag}\Psi\,\big)\Psi{}.
\label{eq:hamiltonian}
\end{equation}
Here $\Psi^{\dag}(\textbf{x})$ and
$\Psi(\textbf{x})$ are the creation and
annihilation operators of the Bose field
in $d$-dimensions. $m$ denotes the mass of
the Bose particles. $v_0>0$ is the
strength of the interaction potential
$V(\textbf{x})= v_0\delta (\textbf{x})$ of
the bosons. Main reason for choosing here
a strictly local interaction is to keep
the number of parameters small, which only
makes a dimensional analysis possible.
However, for low energy phenomena this
approximation is not unreasonable.

It is  convenient to express $v_0$ in
terms of a scattering length $b$ which
follows from dimensional analysis:
\begin{equation}\label{eq:scattering_length}
b=\left(\frac{mv_0}{S_d\hbar^2}\right)^{1/(d-2)}
,\,\,\,\,\,\,d>2.
\end{equation}
$S_d$ is a prefactor with $S_3=4\pi$.  In
$d\le 2$, where
(\ref{eq:scattering_length}) diverges with
$v_0\to 0$, the scattering length behaves
as $b\sim q^{-1}$  where $q$ denotes the
transferred momentum since there is always
a bound state \cite{1}. However,
throughout this paper we will stay in
$d>2$ dimensions.

For $v_0\equiv 0$ the state of the system
is completely described  by the occupation
numbers $n_{\textbf{p}}$ of the single
particle states which are characterized by
their momentum $\textbf{p}$ and energy
$\epsilon_0(\textbf{p})=\textbf{p}^2/(2m)$.
Since the particles of the ideal Bose gas
do not occupy any volume and since there
is no Pauli principle for bosons, in the
ground state all particles can have the
same position.  One conclusion one can
draw from this is that the pressure cannot
depend on the volume \cite{Na_05}. There
are no collective excitations. This
situation is completely changed in the
presence of even a very weak repulsive
interaction. The particles have now a
finite cross-section and occupy a finite
volume. Local density fluctuations will
propagate because of the repulsive forces
between bosons. This behavior results in
the occurrence of a branch of sound waves
i.e. acoustic phonons. The existence of
acoustic phonons in the weakly interacting
system will be the \emph{main assumption}
of the present considerations.

In the \emph{hydrodynamic} limit, i.e. for
frequencies $\omega\ll \tau^{-1}_{relax}$
where $\tau_{relax}$ is a typical
relaxation time of the interacting gas,
the existence of sound waves follows from
the particle number conservation and
Euler's equation (Newton's equation for
liquids) in a linear approximation
\cite{LL_6}. In addition we have to assume
here that these sound waves persist in the
collision-less limit $\omega\gg
\tau^{-1}_{relax}$, i.e. go over into
acoustic phonons. For completeness I
remark that besides of ordinary (first)
sound in superfluids there is second sound
which propagates at $T=0$ with the
velocity $c_2=c/\sqrt{3}$.
 Second sound is a
hydrodynamic mode which does not exist in
the collision-less limit \cite{An_84}.

%%%%%%%%%%%%%%%%%%%%%%%%%%%%%%%%%%%%%%%%%%%%%
\section{T%=U
=0}\label{T=0}
%%%%%%%%%%%%%%%%%%%%%%%%%%%%%%%%%%%%%%%%%%%%%

We begin with the properties of the ground
state. In the non-interacting case all
particle are in the state of lowest energy
$\epsilon_{0}=0$ and
$\langle\Psi^{\dag}\Psi\rangle=n_0=n$
where $n=N/V$. Here $N$ and $V$ denote the
total number of particles and the volume,
respectively. Pressure and the chemical
potential vanish in the ground state.

\medskip

\emph{Phonons}. As discussed already I
will now assume that the excitations above
the ground state can
 still be characterized by the momentum
$\textbf{p}$ (an assumption which
parallels Landaus picture of a Fermi fluid
\cite{An_84}) but that the dispersion
relation $\epsilon(\textbf{p})$ is changed
with respect to $\epsilon_0(\textbf{p})$
such that
\begin{eqnarray}
\label{eq:dispersion}
\epsilon(\textbf{p})&\approx& c|\textbf{p}|,\,\,\,\,\,\,\,\,\, p\lesssim p_0\\
\epsilon(\textbf{p})&\approx&
\frac{\textbf{p}^2}{2m},\,\,\,\,\,\,\,\,\,\,
p> p_0.\label{eq:dispersion2}
\end{eqnarray}
The sound velocity $c$ and the cross-over
momentum $p_0$ have to be determined
self-consistently. Clearly at $p=p_0$
\begin{equation}\label{eq:cross-over}
    c\,p_0\approx
    \frac{p_0^2}{2m},\,\,\,\textrm{i.e.}\,\,\,\,\,\,p_0\approx 2mc.
\end{equation}
I ignore here and in the following all
factors of order unity since their
determination is beyond the accuracy of
our considerations. It is also evident
that for vanishing interaction  the linear
branch (\ref{eq:dispersion}) of the
dispersion relation
 should disappear. As
long as the energy ${p^2}/({2m})$ of the
free bosons is much larger than $v_0n$
(compare (\ref{eq:hamiltonian})) the
effect of the interaction is weak and the
free boson dispersion relation
(\ref{eq:dispersion2}) will essentially
remain unchanged. In the opposite case a
linear dispersion relation
(\ref{eq:dispersion}) is expected. It is
therefore obvious that a second relation
for the determination of $p_0$ is given by
%\begin{figure}
%\centerline{
%\includegraphics[width=0.6\linewidth]{E_p.eps}
%}{\caption{Dispersion relation of
%noninteracting (dashed line) and
%interacting (solid line)
%bosons}\label{Fig:breaks} }
%\end{figure}
\footnote{Some of the formulas resemble
relativistic counterparts , e.g. $\xi$
plays the role of the Compton wave length,
$mc^2$ that of the rest energy etc., but
the crossover from non-relativistic to
relativistic behavior is opposite to the
conventional one: particles with low
momenta behave as relativistic, those with
large momenta as non-relativistic ones.}
\begin{equation}\label{}
\frac{p_0^2}{2m}\approx v_0n\approx mc^2
\end{equation}
 which gives with (\ref{eq:cross-over})
\begin{equation}\label{eq:c}
c={\cal C}_c(d)
    \sqrt{\frac{v_0n}{m}}\sim
    \frac{\hbar}{m\xi_0},
    \,\,\,\,\,\,\,\,\,p_0\sim
    \sqrt{v_0nm}\sim\frac{\hbar}{\xi_0}.
\end{equation}
The relation (\ref{eq:c}) with
 ${\cal C}_c(3)=1$
agrees for $p\ll p_0$ with the well-known
result \cite{Le_57,LL_25.11,An_04}.
Clearly the constant ${\cal C}_c(d)$ (as
well as other constants to come) remains
undetermined in the present approach.
Apparently
 $\xi_0\approx h/p_0$
 plays the role of a fundamental length
scale  of the interacting system besides
of the mean boson distance $a=n^{-1/d}$.
$\xi_0$ is the smallest length scale on
which \emph{collective }behavior is still
seen. In the superfluid phase it
corresponds to the bare correlation or
\emph{healing} length. The assumption of
weak interaction (which will be used
everywhere below) can be written as
\begin{equation}\label{eq:weak_interaction}
   \frac{a}{\xi_0}= \left(\frac{b}{a}\right)^{(d-2)/2}
\ll 1,\,\,\,\,\,\,\,\,a=n^{-1/d}.
\end{equation}

\medskip

\emph{Zero point fluctuations.} The
degrees of freedom which are involved in
the collective phonon excitations have a
non-vanishing zero point energy since they
correspond to an ensemble of harmonic
oscillators. Hence they cannot contribute
to the condensate. The depletion of the
number of particles $N_0=Nn_0$ in the
condensate is therefore given by
\begin{equation}\label{eq:condensate}
   n_{E>0}= n-n_0=
   \sum_{ p<p_0}1={\cal C}_ {n}(d)% N\left(\frac{a}{\xi_0}\right)^d\sim
    n \left(\frac{b}{a}\right)^{\frac{d(d-2)}{2}},
\end{equation}
which  agrees with the exact result
 in $d=3$
dimensions if we choose ${\cal C}_ {
N}(d=3)=\frac{8}{3\sqrt{\pi }}\approx
1.5\,\,\,$ \cite{LL_25.19}. $N_0$ denotes
the number of particles in the condensate.
Thus, contrary to the non-interacting
case, the Heisenberg uncertainty relation
in conjunction with the repulsive
interaction enforces that a finite
fraction of the system is in excited
states.

The  energy of the zero point fluctuations
of the phonons follows analogously from:
\begin{equation}\label{eq:zero_point_fluctuations}
\sum_{p<p_0}\frac{1}{2}{\hbar c
|\textbf{p}|}{} \sim N_{E>0}\,c\,p_0 .
\end{equation}
The total ground state energy $E_0$
consists then of the interaction of the
particles in the condensate and the zero
point fluctuations of the phonons
\begin{eqnarray}\label{eq:ground_state}\nonumber
{E_0}=\frac{1}{2}v_0N_0n_0\,+\,\frac{1}{2}\sum_{p<p_0}{\hbar
c
|\textbf{p}|}{}\,\,=\,\,\,\,\,\,\,\,\,\,\,\,\,\,\\\
\,\,\,\,\,\,\,\,\,\,\,\,
=\frac{v_0N_0^2}{2V_0}\left[ 1+{\cal C}_
{E}(d)\left(\frac{b}{a}\right)^{{d(d-2)/2}}\right].\label{eq:groundstate}
\end{eqnarray}
 With
${\cal C}_
{E}(3)=\frac{128}{15\sqrt{\pi}}\approx
4.8$ the last relation agrees with the
exact result \cite{LL_25.11}. From
(\ref{eq:groundstate}) follows the
chemical potential \cite{LL_25.15}.
\begin{equation}\label{eq:chem_pot}
    \mu=\frac{\partial E_0}{\partial
    N}=v_0n_0\left[
1+\frac{d+2}{4}{\cal C}_
{E}(d)\left(\frac{b}{a}\right)^{{d(d-2)/2}}\right].
\end{equation}
  The compressibility
$\kappa=\partial n/\partial\mu$ is then
\begin{equation}\label{eq:compressibility}
\kappa^{-1}=v_0\left[1+\frac{d(d+2)}{8}{\cal
C}_
{E}\left(\frac{b}{a}\right)^{{d(d-2)/2}}\right].
\end{equation}
Finally we calculate the pressure
\begin{equation}\label{eq:pressure}
    p=-\frac{\partial
    E_0}{\partial V}\approx
    \frac{1}{2}v_0n_0^2\sim \frac
    {\hbar^2}{ma^{d+2}}\left(\frac{b}{a}\right)^{d-2}
\end{equation}
It is instructive to express the results
(\ref{eq:condensate}) -
(\ref{eq:pressure}) in terms of the
healing length
(\ref{eq:weak_interaction}).

%%%%%%%%%%%%%%%%%%%%%%%%%%%%%%%%%%%%%%%%%%%%%%%%%%%%%%%%%%%%%%%%%%%%%%%%%%
\section {T $>$ 0}
%%%%%%%%%%%%%%%%%%%%%%%%%%%%%%%%%%%%%%%%%%%%%%%%%%%%%%%%%%%%%%%%%%%%%%%%%%%

\emph{Ideal gas.} We begin again with a a
brief review of the situation in the
absence of any interaction \cite{Na_05}.
Then at temperatures below the Bose
condensation  temperature of the ideal
gas,
\begin{equation}\label{}
    T_{c,0}={\cal C}_{T_c}\frac{\hbar^2}{ma^2},
\end{equation}
where ${\cal C}_{T_c}=3,31$ \cite{LL_5},
of the order $
N\big({a}/{\lambda_{T,2}}\big)^d$
particles are outside of the condensate.
Here $\lambda_{T,2}=\hbar/\sqrt{mT}$
denotes the thermal de Broglie wave length
of the
 non-relativistic Bose particles. The pressure follows as
$p\sim T/\lambda_{T,2}^d$ etc., see
\cite{Na_05}.

\medskip
%%%%%%%%%%%%%%%%%%%%%%%%%%%%%%%%%%%%%%%%%%%%%%%%%%%%%%%%%%%%%%
%\vspace{0.5cm}
\emph{Weak interaction, low
$T$}. For non-zero interaction
\begin{equation}\label{}
    v_0n_0\approx mc^2\approx \frac{\hbar^2}{m\xi^2}\sim T_{c,0}\left(\frac{b}{a}\right)^{d-2}
 \ll T_{c,0}
\end{equation}
sets a second energy scale. As long as
$T\ll mc^2$ the relevant part of the
excitation spectrum is essentially linear
with a thermal de Broglie wave length
\cite{Na_05}
\begin{equation}\label{}
\lambda_{T,1}=\frac{\hbar c}{T}\sim
\,\frac{T_{c,0}}{T}\frac{a^2}{\xi_0}\sim
a\,\frac{T_{c,0}}{T}\left(\frac{b}{a}\right)^{(d-2)/2}.
\end{equation}
Since there is no phonon number
conservation, $\lambda_{T,1}$ has the
meaning of  the average spacing between
the phonons. Since phonons behave as
non-interacting relativistic particle we
obtain their entropy $S$ by using the
relation \cite{Na_05} $S={\cal
C}_S(d)V/\lambda^d_{T,1}$ for relativistic
particles of energy $T$. The exact
expression is  ${\cal
C}_S(3)=4\pi^2/90\approx 0.44\,\,\,$
\cite{LL_63.13}. Thus their energy $E$ is
given by \cite{Na_05} $E_0 =
\frac{d}{d+1}{\cal C}_S(d)
V\frac{T}{\lambda_{T,1}^d}$  and hence we
get for the total free energy in the range
$T\ll mc^2\sim T_{c,0}(b/a)^{d-2}\,\,\,\,$
\cite{LL_22.4}
\begin{equation}
{E} =\frac{n_0^2v_0V}{2} \left[
 1 +
 \left(\frac{b}{a}\right)^{\frac{d(d-2)}{2}}
 \Big[{\cal C}_{E}+{\cal C}_S
\left(\frac{T}{mc^2}\right)^{d+1}\Big ]
\right],\,\,\,\label{eq:free_energy}
\end{equation}
in agreement with the so-called Popov
approximation \cite{LL_25.11,Po_83}. The
finite temperature corrections are small,
as has to be expected. % and in agreement.
Specific heat, internal energy, pressure
etc. follow from that.

\medskip

\emph{Superfluid density}. At zero
temperature the whole Bose liquid is
superfluid. At finite $T$ however, part of
it has normal viscous  properties. To
obtain the result for the normal and
superfluid density, $n_n$ and $n_s$,
respectively, we follow the ideas of
Landau and Khalatnikov \cite{Kh_65}: We
decompose the particle current density in
a normal and a superfluid one
\begin{equation}\label{eq:current}
    \textbf{j}=n_s\textbf{v}_s +n_n\textbf{v}_n,\,\,
    \,\,\,\,n_s+n_n=n.
\end{equation}
Here $\textbf{v}_s$ and $\textbf{v}_n$
denote the velocity of the normal and the
superfluid  component, respectively, in
the laboratory frame. In the frame in
which the superfluid is at rest the
current is
\begin{equation}\label{eq:current_2}
    \textbf{j}_0=\textbf{j}-\textbf{v}_sn=n_n(\textbf{v}_n-\textbf{v}_s).
\end{equation}
$m\textbf{j}_0$ denotes the mass current
density (i.e. the momentum {density}) of
all particles which do \emph{not} belong
to the superfluid. This is the net
momentum {density} resulting from all
phonon degrees of freedom. It can be
written as
\begin{equation}\label{eq:normal_current}
   n_n\left(\textbf{v}_n-\textbf{v}_s\right) =\hbar^{-d}\int
   d^dp\,\frac{\textbf{p}}{m}n_{ph}\left(cp+\textbf{p}(\textbf{v}_s-\textbf{v}_n)\right)
\end{equation}
where $n_{ph}$ denote the phonon
distribution function (which we don't have
to know!). We expand the r.h.s.  up to
linear terms in $\delta
\textbf{v}=\textbf{v}_s-\textbf{v}_n$,
multiply the whole equation by $\delta
\textbf{v}$ and perform the angular
average. This gives
\begin{eqnarray}\label{eq_n_n}
\nonumber
  n_n &=& \frac{1}{md}\int
  d^dp\,p^2\,\frac{\partial\,
    n_{ph}}{\partial cp}\,
    =\frac{d+1}{d}\frac{1}{mc^2}\int
    d^dp\,
    cp\,n_{ph}(p)\\
   &=&n\,{\cal C}_S(d)
   \left(\frac{T}{mc^2}\right)^{d+1}
   \left(\frac{b}{a}\right)^{\frac{d(d-2)}{2}}
 =
n\frac{S}{N}\frac{T}{mc^2}.
\end{eqnarray}
In the last steps we integrated by parts
and expressed the resulting integral in
terms of the energy and entropy,
respectively. It is remarkable that the
last relation is independent of dimension.
Clearly the superfluid component has zero
entropy. .

\bigskip

 \textsc{\textbf{Acknowledgment}}

It is a pleasure to acknowledge helpful
advice on the preparation of this paper by
Jens O.  Andersen, Sergey Artemenko,
Andreas Engel, Axel Pelster, Aleksandra
Petkovits, Zoran Ristivojevic, Friedmar
Sch\"utze and Matthias Vojta.

%%%%%%%%%%%%%%%%%%%%%%%%%%%%%%%%%%%%%%%%%%%%%%%%%%%%%%%%%%%%%%%%%%%%
%%%%%%%

%%%%%%%%%%%%%%%%%%%%%%%%%%%%%%%%%%%%%%%%%%%%%%%%%%%%%%%%%%%%%%%%%%%%%%%%


\begin{thebibliography}{99}
\bibitem{Na_05} T. Nattermann, \emph{''A scaling approach to ideal quantum gases'}'
Am. J. Phys \textbf{73}, 349 (2005).

\bibitem{1}  In three dimensions
(\ref{eq:scattering_length}) agrees with
the textbook result in the limit of small
wave vectors. For scattering in two
dimensions see

D.S. Petrov and G.V. Shlyapnikov,
\emph{''Interatomic collisions in tightly
confined Bos gas'',} Phys. Rev. A
\textbf{64}, 12706 (2001).

\bibitem{LL_6} L.D. Landau and
E.M. Lifshitz, \emph{''Fluid Dynamics''},
2nd edition, Pergamon Press Oxford,
\S\textbf{64} (1987).

\bibitem{An_84} P.W. Anderson, \emph{''Basic
Notions of Condensed Matter Physics''},
Benjamin/Cummings Publishing Company,
Menlo Park, California (1984).


\bibitem{Le_57} Most of the results for the
weakly interacting Bose gas were
originally obtained by T.D. Lee, K. Huang
and C.N. Yang in \emph{''Eigenvalues and
Eigenfunctions of a Bose System of Hard
Spheres and Its Low-Temperature
Properties'}', Phys. Rev. \textbf{106},
1135-46 (1957).

\bibitem{LL_25.11} We refer here to the widespread
textbook of L.D. Landau and E.M. Lifshitz,
\emph{''Statistical Physics, Part 2''},
second edition, Pergamon Press Oxford
(1981), eq. (25.11).

\bibitem{An_04} More results on the weakly interacting
Bose gas can be found in J.O. Andersen,
\emph{''The Weakly Interacting Bose
Gas''}, Rev. Mod. Phys. \textbf{76},
599-639 (2004).

\bibitem{LL_25.19} Reference \cite{LL_25.11}, eq.
(25.19).

\bibitem{LL_25.14} Reference \cite{LL_25.11}, eq.
(25.14).
\bibitem{LL_25.15} Reference \cite{LL_25.11}, eq.
(25.15).


\bibitem{LL_5} L.D. Landau and E.M. Lifshitz,
\emph{''Statistical Physics, Part 1''},
third edition, Pergamon Press Oxford
(1980), eq.(62.2).


\bibitem{LL_63.13} Reference \cite{LL_5},
eq. (63.13).

\bibitem{LL_25.11_27.10} Reference
\cite{LL_25.11}, eq. (27.10).

\bibitem{LL_22.4} Reference \cite{LL_25.11}, eq.
(22.4).




\bibitem{Po_83} V.N. Popov, \emph{''Functional
Integrals in Quantum Field Theory and
Statistical Physics''}, Reidel. Dortrecht
(1983)

\bibitem{Kh_65} see in particular I.M. Khalatnikov \emph{''An
Introduction to the Theory of
Superfluidity''}, W.A. Benjamin Inc, New
York, Amsterdam (1965).

\end{thebibliography}
\end{document}